\documentstyle[12pt]{article}

\begin{document}

\hsize 14.5 cm 

\centerline {\Large International Centre for Social Complexity, Econophysics}
\centerline {\Large \& Sociophysics Studies: A Proposal}

\vskip 1 cm

\centerline {\large Bikas K. Chakrabarti}
\medskip
\centerline {Saha Institute of Nuclear Physics, Kolkata 700064, India}
\centerline {S. N. Bose National Centre for Basic Sciences, Kolkata 700106, India}
\centerline {Economic Research Unit, Indian Statistical Institute, Kolkata 700108, India}

\vskip 2 cm

\noindent {\bf Abstract:} In the concluding session of the Joint 
International Conference titled `Econophys-2017 \& Asia Pacific 
Econophysics  Conference (APEC)-2017', held in Jawaharlal
Nehru University and Delhi University during  November 
15-18, 2017, a brief version of this Proposal was
presented. Several important and enthusiastic comments were
received from the participants. This note is based on
these comments and discussions. 

\vskip 1 cm

\leftline {\bf 1. Introduction}

\medskip

\noindent More than twenty years have passed since the formal
coining of the term and hence the launch of
econophysics as a research topic (since 1995; see
the entry by Barkley Rosser on Econophysics in `The
New Palgrave Dictionary of Economics' [1]: {\it ``Econophyics:
According to Bikas Chakrabarti (2005), the term
`econophysics' was neologized in 1995 at the second
Statphys-Kolkata conference in Kolkata (formerly
Calcutta, India) by the physicist H. Eugene Stanley ..."}).
Soon, econophysics had been assigned the Physics
and Astronomy Classification Scheme (PACS) number 89.65Gh
by the American Institute of Physics. According to
Google Scholar, typically today more than thousand papers
and documents, containing the term `econophysics', are
published each year (many more  research papers are, in fact,
published today on the topic without ever calling it
econophysics) in almost all physics journals covering
statistical physics, general science journals and a
few economics journals. More than fifteen books on
econophysics (with the word econophysics in the title
of the book), including some textbooks and monographs written by
pioneers and active researchers in the field, have already
been published by Cambridge University Press,  Oxford
University Press, Springer and Wiley. Many more edited
books and conference proceedings are published (search of `econophysics'
titles in the `amazon.com:books' today gives more than
140 entries; with some double counting of course!). 
Similar has been the story for
`sociophysics'.

Regular interactions and collaborations between the
communities of natural scientists and social scientists,
however,  are rare even today! Though, as mentioed already,
interdisciplinary research papers on econophysics and
sociophysics are regularly being published at a steady and
healthy rate, and a number of universities (including
Universities of Bern, Leiden, London, Paris and Tufts University) are
offering the interdisciplinary courses on econophysics
and sociophysics, not many clearly designated professor
positions, or other faculty positions for that matter,
are available yet (except for econophysics in Universities
of Leiden and London). Neither there are  designated
institutions on these interdisciplinary fields, nor
separate departments or centres of studies for instance.
We note however, happily in passing, a recently published
highly acclaimed (`landmark' and `masterful') economics
book [2] by Martin Shubik (Seymour Knox Professor
Emeritus  of
Mathematical Institutional Economics,  Yale
University) and Eric Smith (Santa Fe Institute) discusses
extensively on econophysics approaches and in general on
the potential of interdisciplinary researches inspired by
the developments in natural sciences. Indeed, this massive
580-page book can also serve as an outstanding `white-paper'
document in favor of our intended Proposal.

Though the inter-disciplinary interactions have not 
grown much, some sure signs of positive impact for
the research achievements in econophysics and sociophysics
have been documented in the literature. The precise 
characterizations of stock market fluctuations by
Mantegna and Stanley [3] has already made a decisive mark in
financial economics and
all the related subjects (with more than 4000 citations
already for the book [3];
Google scholar). In the section on `The position of 
econophysics in the disciplinary space' in the book `Econophysics and 
Financial Economics' [4], the authors write (pp. 83, 178): {\it
``To analyze the 
position of econophysics in the disciplinary space, the most 
influential authors in econophysics were identified. Then 
their papers in the literature were tracked by using the 
Web of Science database of Thomson-Reuters ... The sample 
is composed of Eugene Stanley, Rosario Mantegna, Joseph 
McCauley, Jean Philippe Bouchaud, Mauro Gallegati, Benoit 
Mandelbrot, Didier Sornette, Thomas Lux, Bikas Chakrabarti, 
and Doyne Farmer."} The book [2] by Shubik and Smith noted (pp. 75-76)
 that while simple 
kinetic exchange market model (see e.g., [5])  leads to
exponentially decaying distributions,  {\it``it was shown in 
[6]  that uniform saving propensity of the agents 
constrains the entropy maximizing dynamics in such a way 
that the distribution becomes gamma-like, while (quenched) 
nonuniform saving propensity of the agents leads to a 
steady state distribution with a Pareto-like power-law 
tail [7]. 
A detailed discussions of such steady state distributions 
for these and related kinetic exchange models is provided 
in  [8]"}.
Shubik and Smith [2]  also noted the
important contributions by physicists in the study of
multi-agent iterative (and collective) learning game  models
for efficient resource sharing ([9] for 
binary choice iterative learning games and [10] 
for multi-choice iterative learning  
games\footnote{\label{*}Important developments have taken place in 
such many-player, multi-choice iterative learning games for limited 
resource utilizations, since publication of The Kolkata Paise 
Restaurant Problem \& Resource Untilization, A. S. Chakrabarti, 
B. K. Chakrabarti, A. Chatterjee and M. Mitra,  
Physica A, {\bf 388}, pp.  2420-2426 (2009). For applications to quantum 
cryptography physics, computer job scheduling, on-line car hire, etc., see 
e.g., Strategies in Symmetric Quantum Kolkata Restaurant Problem,
P. Sharif \&  H. Heydari,  Quantum
Theory: Reconsideration of Foundations 6: AIP Conf. Proc. {\bf 1508},
pp. 492-496 (2012); Econophysics of the Kolkata Restaurant
Problem \& Related Games; B. K. Chakrabarti, A. Chatterjee, A. Ghosh, S.
Mukherjee \& B. Tamir, Springer (2017); Econophysics \& the Kolkata 
Paise Restaurant Problem: More is Different, B. Tamir, Science \& Culture,
{\bf 84}, pp. 37-47 (2018); The Vehicle for Hire Problem: A Generalized
Kolkata Paise Restaurant Problem, L. Martin \& P. Karaenke, 
https://mediatum.ub.tum.de/doc/1437330/1437330.pdf  (2018);  Kolkata 
Paise Restaurant Game for Resource Allocation in the Internet 
of Things, T. Park \& W Saad, IEEE Xplore, DOI: 10.1109/ACSSC.2017.8335666,
https://ieeexplore.ieee.org/abstract/document/8335666/ (2018)  }). 
This book [2] 
also discusses in details
on the impact of the  pioneering work by physicist 
Per Bak and collaborators in the context of
self-organizing dynamics of 
complex  markets. The Econophysics course offered  by Diego 
Garlaschelli in the Physics department of the Leiden 
University, where the first economics Nobel laureate 
(statistical physicist Jan Tinbergen)
came from, follows exclusively the book  
`Econophysics: An Introduction' [11] 
since its inception in 2011 (see e.g., [12] for the 2017-2018
and 2018-2019 e-prospectuses).
Discussions on some more impact of econophysics [3, 4, 13, 14]
and sociophysics [15-18] researches will be continued 
later.

\vskip 1 cm

\leftline {\bf 2. Proposal in Brief and Some Earlier Attempts}
\medskip

\noindent In view of all these, it seems it is time to try for 
an international centre for interdisciplinary studies 
on complexity in social and natural sciences; 
specifically on econophysics and sociophysics. The model
 of the Abdus Salam International Centre for Theoretical 
Physics (ICTP), Trieste (funded by UNESCO and IAEA), 
could surely be helpful to guide us here. We are 
contemplating, if an ICTP-type interdisciplinary research 
institute could be initiated for researches on econophysics 
and sociophysics (see also  [19]).

We note that Dirk Helbing (ETH, Zurich) and colleagues
have been trying for an European Union funded `Complex
Techno-Socio-Economic Analysis Center' or `Economic and
Social Observatory' for the last six years (see Ref. [20]
containing the White Papers arguing for the proposed
centre). We are also aware that Indian Statistical
Institute had taken a decision to initiate a similar
centre in India (see  `Concluding Remarks' in [21]). Siew
Ann Cheong (Nanyang Technological University, Singapore)
had tried for a similar Asian Centre in Singapore [22].
In view of some recent enthusiasms at the Japan-India
Heads of States or Prime Minister level, and signing of
various agreements (predominantly for business deals,
infrastructure development, technical science and also
cultural exchanges) by them, possibility of an Indo-Japan
Centre for studies on Complex Systems is also being
explored, including the possibility of a centre in Tokyo
with private support [23]. There are several other similar
initiatives (e.g., [24]).

These proposals are, or had been, for regular research centres on
such interdisciplinary fields, where regular researchers
are expected to investigate such systems.
In view of the extreme interdisciplinary nature of
econophysics and sociophysics, such efforts may be
complemented by another visiting centre model.

Unlike the above-mentioned kind of intended centres,
this proposed centre may be just a visiting centre where
natural and social scientists from different universities
and institutions of the world can meet for extended
periods to discuss and interact on various
interdisciplinary issues and collaborate for such
researches, following the original ICTP model. Here,
as in ICTP, apart from a few (say, about ten to start-with)
promising young researchers on econophysics and sociophysics
as permanent faculty who will continue active research
and active visiting scientist programs (in physics,
economics and sociology) etc. can be pursued, The faculty
members, in consultation with the advisers from
different countries, can choose the invited visitors and
workshops or courses, on economics and sociological
complexity issues, can be organized on a regular basis
(as for basic theoretical sciences in ICTP or in Newton
Centre, Cambridge, etc.). In two short communications [25],
Martin Shubik (Yale University, New Haven) supported the
idea very enthusiastically  and encouraged us with some
very precise suggestions. He also noted that such a centre
can play a much more inclusive role for the whole world
(as is being done by the ICTP), compared to what the
Santa Fe Institute has been successful to do for the US.
Gene Stanley (Boston University, Massachusetts) supported
enthusiastically such a proposal ({\it ``... you already
thought of all the ideas I might have had ... I will
continue to think ... congratulations on your ambitious
idea ... "} [26]).

\vskip 1 cm

\leftline {\bf 3. Some Responses Received From the Participants}

\medskip

\noindent After my brief  presentation of this proposal in the
Concluding session of our Conferences, there were several
appreciative comments made by the participants and a
number of precise suggestions mailed to me later by
many participants including Frederic Abergel (Centrale
Supelec, Chatenay-Malabry Cedex), Bruce Boghosian  (Tufts
University, Massachusetts), Anirban Chakraborti 
(Jawaharlal Nehru University, Delhi), Siew Ann Cheong (Nanyang Technological 
University, Singapore), Acep Purqon (Institute for Technology, 
Bandung) and Irena Vodenska (Boston University, 
Massachusetts). 
I append below parts of a few detailed
comments, summarizing  the past achievements and some 
suggestions for possible structural organisation, received  
from  them:

\medskip

\noindent {\bf A)} Regarding the ``{\it discoveries of important economics and 
finance phenomena that were unknown to economists and financial 
economists before, the following few  come to my mind:

\medskip

\noindent a)  The distribution of wealth and income.  While Pareto was 
the first to
examine the tail end of the wealth distribution, and 
found it to be a
power law, little was known and understood about the full 
distribution
until you and Victor Yakovenko came along, to (i) examine empirical
distributions of wealth and income [27],
and (ii) build
kinetic
theory/agent-based models to show that the full distribution is an
exponential distribution crossing over to a power-law tail [6, 28]  
 and this arise
because for rich people, they can gain from return on investment or
through interests generated by savings, whereas the rest of us, repeated
random exchange of income/wealth shape the exponential part of the
distribution. During Econophys APEC 2017, we heard Bruce talking about his
further results showing that if wealth is inadequately redistributed
through taxation, oligarchs emerge, leading to the most extreme form of
wealth inequality that we can possibly imagine [29, 30].

\medskip

\noindent b)     Home prices and property bubbles. Following your lead, and more
recently the work by Ohnishi et al. [31],
 my students
and I have started looking
into the distribution of home prices around various markets.
Interestingly, the equilibrium distribution of home prices is similar to
the income/wealth distribution, consisting of an exponential body and a
power-law tail [32].
We see this in
Singapore, Hong Kong, Taiwan, UK, and Japan
so far, and believe this result is universal. We also found that in bubble
years, the home price distribution develop dragon kings, which are strong
positive deviations from the equilibrium distribution. We have evidence to
suggest that such dragon kings are the results of speculation, but have
yet to test regulations that can help defuse them in agent based models
that we are currently building. More alarmingly, we have seen from the
historical home price data of London and Tokyo that their distributions
once contained an exponential body, but after experiencing a couple of
property bubbles, have become asymptotic power laws with no exponential
body. This is another manifestation of economic inequality, in that for
cities like London and Tokyo, homes are priced out of the reach of the
middle class. From the historical data for UK, we see this trend repeating
itself for cities like Birmingham and Manchester. This calls for action on
the part of government, but they cannot act until we understand the
processes that drive this trend.

\medskip

\noindent c)     Louis Bachelier was the first to propose that stock returns
perform
Brownian motion, and laid the mathematical foundation for finance.
However, for a long time, it has not occurred to financial economists to
check the validity of Bachelier¹s assumptions. Benoit Mandelbrot did so in
1967, and found that the tail of the return distribution is a power law [33].
Rosario and Gene then demonstrated more convincingly using a large data
set of returns for the S\&P 500 in their 1995 Nature paper that the return
distributions for different time horizons follow a scaling form, and this
scaling form can be fitted better to a Levy distribution than to a
Gaussian distribution [34]. 
Since then, many different agent based models have
been developed to explain the emergence of fat tails in the return
distribution. More recently, Hideki and Misako Takayasu examined
high-frequency order data, and demonstrated convincingly that stock price
is an invisible Œparticle¹ performing stochastic motion as a result of it
being bombarded on either side by bid and ask orders [35].
 For regular Brownian
motion, this noise is uncorrelated in time, and therefore we end up with
long autocorrelations in the velocity of the Brownian particle. For stock
returns, we know from many previous works that they are nearly
uncorrelated in time. The Takayasus explained that this is the consequence
of the noise being strongly correlated in time, pointing to what they
observe in the order book data. This duality is surprising!

\medskip

\noindent d)     Economists Ricardo Haussmann and 
Cesar Hidalgo became world famous
for publishing their Atlas of Economic Complexity [36],
 visualising the network of
international trade over time. Not convinced that the economists have
extracted the most important insights from the data, Luciano Pietronero
went in to the data set to plot the economic performances of countries on
a two-dimensional plot, with capabilities on the x-axis, and GDP on the 
y-axis [37].
Luciano found that he could classify countries into undeveloped,
developing, and developed economies by where they appear on the plot.
Undeveloped countries are problematic, and are mostly African, because
their GDPs are low, and their capabilities are also low. These countries
can potentially be stuck in a poverty trap, because they earn so little
that they cannot reinvest into their education system to increase their
capabilities. Developing countries like China, India, and Vietnam are
countries that have in the past invested heavily into education and are
therefore ranked highly in terms of their capabilities. China has already
started to benefit from its past investment, to see a steady rise in its
GDP. India can be seen to be following suit, and Vietnam will likely take
off soon. When Luciano produced such plots using data from different
years, he found that the developing countries are in a region where
economic trajectories are fairly deterministic, and therefore we can have
confidence in the economic futures of India and Vietnam, for example. On
the other hand, the undeveloped countries are in a region of the plot
where economic trajectories appear to be chaotic and turbulent, where
countries can experience periods of enhanced GDP because of exploitation
of resources (like Brazil), but can also fall from grace just as quickly
because of political turmoil.

In creating this list, I am leaving out interesting results obtained by
people working on urban complexity, because they rarely attend
econophysics conferences. Besides the most important scaling work done by
Geoffrey West and Luis Bettencourt, showing that there are urban variables
that scale sub linearly with the size of cities, and other urban variables
(like GDP, patents, crime, etc.) that scale super linearly with size [38].
Hyejin
Youn and her collaborators have also found that cities are not equally
diverse in terms of job opportunities [39].
Small cities tend to have fewer
types of jobs, and more people working on the same type of jobs. Large
cities tend to have more types of jobs, and fewer people working on each
type of job. More importantly, they have discovered that wealth is
unequally concentrated in large cities, and that large cities tend to have
a better educated populace, and because of this, is more resilient against
the ongoing economic restructuring due to automation.

\medskip

Finally, besides telling success stories, we also 
need to frame a few key questions
that we hope the international centre can address. 
Here, we should be ambitious, and
go for questions that individual investigators, or 
even individual universities
would not have the capability, resource, or correct 
composition of different experts
to address."}

\bigskip 

\noindent {\bf B)}  In this connection, it may be worthy to
note that  {\it ``the German Physical Society has a 
working group on Physics of
Socio-Economic
Systems since 2009 (see e.g., [40]: ... This dedicated 
scientific community is rapidly growing and involves, 
besides sociologist and economists, also physicists, 
mathematicians, computers scientists, biologists, 
engineers, and the communities working on complex 
systems and operations research ...). Apart from 
supporting researches and recognising
regularly active young researchers (with `Young Scientist Award 
for Socio- and Econophysics') in such interdisciplinary 
fields, they organise many conferences within Germany with
participants from all over Europe."}.

\bigskip

\noindent {\bf C)} Regarding a possible financial structure, 
``{\it I note, following Shubik, we want to raise funds for 
it to be endowed in perpetuity and cost of the regular 
activities can be met from the (fixed deposit) interests.
As we discussed in Delhi after the Conference, this is not easy
but I am hopeful. Also, I agree with Shubik, it is worth trying. 
 ... Presumably, to begin with, the founding faculty members would
need only a fraction of their salary, and the bulk of the
interest money could be used for postdocs, graduate student support,
visitor travel, etc.
For a different institutional model, have a look at
the web page of ICERM at Brown University
(https://icerm.brown.edu/home/index.php ).  From our
conversations in New Delhi, I understand that you would
like to see a more extensive and inclusive model for
this purpose, located somewhere in Eurasia, and I am
very supportive of this idea.

To raise funds for this kind of thing, it will be
necessary to create a clear proposal that addresses -- at
the very minimum -- the following items:
\medskip

\noindent a) First, we need a list of names and bios of
international faculty who would be willing lend their names
to such a Centre.  In fact, it would be better to partition
this list into categories:  Some more senior faculty
with administrative experience could serve on an Advisory
Board. Other faculty would be willing to visit the Centre
from time to time, and perhaps organize conferences there.
Some would send their graduate students during the summer,
etc.

\medskip

\noindent b) Second, we need a clear business model for the Centre,
along with a governance model and sample budget.  Again, we
might learn from the models of ICERM, ICTP and SFI, but we
probably want something that is unique to what we have in
mind.
\medskip

\noindent c) Third, we need a list of benefits from this proposed
Centre that would accrue to the hosting institution and the
hosting country."}

\vskip 1 cm

\noindent {\bf 4. Concluding Remarks}

\medskip

\noindent We think, it is an appropriate time to initiate such a
project for the healthy  growth of this `Fusion of Natural
and Social Sciences', through active dialogue among the
students and experts from different disciplines (e.g., physics, computer
science, mathematics, economics and sciology), engaged in
researches in their respective disciplines and institutions, from
all over the world. We find, both the experts in the
related disciplines as well as the researchers already
initiated in such interdisciplinary researchers have
deep feelings about the urgent need for such a Centre, where
short and long term visits would be possible and enable them to participate in
interdisciplinary schools, workshops, and research
collaborations.

\vskip 1 cm

\noindent {\bf Acknowledgement:} I am grateful to Yuji Aruka, Arnab 
Chatterjee, Asim Ghosh, Taisei Kaijozi, for several interactive  
discussions at an
earlier stage. Comments on a  draft of this Note from Abhik Basu,
Soumyajyoti Biswas, Indrani Bose,  Anirban Chakraborti and Parongama Sen 
 are also gratefully acknowledged.
 I am indebted, in particular, to Bruce 
Boghosian and Siew Ann Cheong for their recently mailed 
 extremely supportive comments 
and detailed suggestions which,
with their kind permission, have been partly included here. I am
thankful to J. C. Bose National Fellowship (DST, Govt. India) for 
support.

\vskip 1.5 cm

\noindent {\bf References}

\medskip

\noindent [1] J. Barkley Rosser Jr., Econophysics, 
in  {\it The New Palgrave
Dictionary of Economics}, S. N. Durlauf and L. E. Blume, eds.,
Palgrave  Macmillan, New York,  
 vol. 2, pp. 729-732 (2008)

\medskip

\noindent [2] M. Shubik and E. Smith, {\it The Guidance of an Enterprise
Economy}, MIT Press, Cambridge, Massachusetts (2016)

\medskip

\noindent [3] R. N. Mantegna and H. E. Stanley, {\it An Introduction to
Econophysics}, Cambridge University Press, Cambridge  (2000) 

\medskip

\noindent [4] F. Jovanovic and C. Schinckus, {\it Econophysics and
Financial Economics}, Oxford University Press, Oxford (2017)

\medskip

\noindent [5] V. M. Yakovenko and J.  Barkley Rosser, Statistical 
mechanics of money,
Reviews of Modern Physics, {\bf 81}, 1703-1725 (2009)

\medskip
\noindent [6] A. Chakraborti and B. K. Chakrabarti, 
Statistical mechanics of money: How saving propensity affects 
its distribution, European Physical Journal B,
{\bf 17}, pp. 167-170 (2000)
\medskip

\noindent [7] A. Chatterjee, B. K. Chakrabarti and 
S. S. Manna, Pareto law in a kinetic model of market with
random saving propensity, Physica A {\bf 335},
pp. 155-163 (2004)
\medskip

\noindent [8] B. K. Chakrabarti, A. Chakraborti, 
S. R. Chakravarty and A. Chatterjee, {\it Econophysics of
Income and Wealth Distributions},
Cambridge University Press, Cambridge (2013)
\medskip

\noindent [9] D. Challet, M. Marsili and Y.-C. Zhang, {\it
 Minority Games}, Oxford University Press, Oxford (2004)
\medskip

\noindent [10] A. Chakraborti, D. Challet, A. Chatterjee,
M. Marsili, Y.-C. Zhang and B. K. Chakrabarti, Physics 
Reports, {\bf 552}, pp. 1-26 (2015)

\medskip

\noindent [11] S. Sinha, A. Chatterjee, A. Chakraborti and
B. K. Chakrabarti, {\it Econophysics: An Introduction}, Wiley, Berlin 
(2010)
\medskip

\noindent [12] Econophysics, 2017-2018 \& 2018-2019 :e-Prospectuses, 
Leiden University

\noindent [https://studiegids.leidenuniv.nl/en/courses/show/69415/econofysica]

\noindent [https://studiegids.leidenuniv.nl/courses/show/81929/econofysica]

\medskip

\noindent [13] P. Richmond, J. Mimkes and S. Hutzler, {\it 
Econophysics and Physical Economics}, Oxford University Press, Oxford  (2013)
\medskip

\noindent [14] F. Slanina, {\it Essentials of Econophysics Modelling},
Oxford University Press, Oxford (2014)

\medskip

\noindent [15] C. Castellano, S. Fortunato and V.
Loreto, Statistical physics of social dynamics, 
Reviews of Modern Physics, {\bf 81}, 591-646 (2009) 

\medskip

\noindent [16] D. Stauffer,  Opinion Dynamics and 
Sociophysics, in {\it Enclypedia 
of Complexity and Systems Science}, Ed. R. A. Meyers, Springer, 
N. Y., pp. 6380-6388 (2009)

\medskip

\noindent [17] S. Galam, {\it Sociophysics}, Springer, N. Y. (2012)

\medskip

\noindent [18] P. Sen and B. K. Chakrabarti, {\it 
Sociophysics: An Introduction},
Oxford University Press, Oxford (2013) 
\medskip

\noindent [19] A. Chatterjee, A. Ghosh and B. K. Chakrabarti, in {\it Economic
Foundations for Social Complexity Science: Theory,
Sentiments, and Empirical Laws}, Eds. A. Kirman and Y. Aruka,
Springer, Tokyo, pp. 51-65 (2017)

\noindent [arXiv: https://arxiv.org/pdf/1611.00723.pdf ].

\medskip

\noindent [20] D. Helbing and S. Balietti, Complex techno-socio 
economics, European Physical Journal: Special Topics, {\bf 195}: pp. 1-136
(2011); see also, Comments by B. K. Chakrabarti, ibid. pp. 145-146,
and the comments by other reviewers for the Proposal and the 
responses of the proposers, ibid. pp. 137-186

\medskip

\noindent [21] A. Ghosh, Econophysics research in India in the
last two decades, IIM Kozhikode Society \& Management Review,
Sage Publications,
{\bf 2}(2) pp.135-146 (2013)

\medskip

\noindent [22] S. A. Cheong, Private communications (2013)

\medskip

\noindent [23] Y. Aruka and T. Kaizoji, Private communications (2016-2017)

\medskip

\noindent [24] T. Di Matteo, Private communications (2016-2017)

\noindent [See also: https://econophysicsnetwork.kcl.ac.uk/ ]

\medskip 

\noindent [25] M. Shubik, Private communications (November-December, 2016)

\medskip

\noindent [26] H. E. Stanley, Private communications (November, 2017)

\medskip

\noindent [27] A. Dragulescu and V. M. Yakovenko,
Exponential and power-law probability distributions of wealth and
income in the United Kingdom and the United States, Physica A,
 {\bf 299}, pp. 213-221 (2001)

\medskip

\noindent [28] A. Dragulescu
and V. M. Yakovenko,  Statistical mechanics of money, European
Physical Journal B, {\bf 17},
pp. 723-729 (2000)

\medskip

\noindent [29]  A. Devitt-Lee, H. Wang, J. Li and B.M. Boghosian, 
A non-standard description of
wealth concentration in large-scale economies, SIAM Journal
of Applied Mathematics, {\bf 78}(2), pp. 996-1008 (2018)

\medskip

\noindent [30] B. M. Boghosian, A. Devitt-Lee, M
Johnson, J. Li, J. A.  Marcq  and H. Wang,  Oligarchy as a
phase transition: The effect of wealth-attained advantage in a
Fokker-Planck description of asset exchange. Physica A,
 {\bf 476}, pp. 15-37 (2017)

\medskip

\noindent [31] T. Ohnishi, T.  Mizuno, C.  Shimizu
and T. Watanabe,  Power laws in real estate prices during bubble
periods. International Journal of Modern Physics: Conference Series,
World Scientific,  {\bf 16}, pp. 61-81 (2012)

\medskip

\noindent [32] D. J. Tay, C.-I. Chou, S.-P. Li, S. Y. Tee and S. A.
Cheong, Bubbles
are departures from equilibrium housing markets: evidence from Singapore
and Taiwan, PLoS ONE, {\bf 11}(11) (November 3, 2016)

\medskip

\noindent [33] B. Mandelbrot,  The variation of some other 
speculative prices, The Journal of Business, {\bf 40},  pp. 393-413 (1967)

\medskip

\noindent [34] R. N. Mantegna and H. E. Stanley,  Scaling
behaviour in the dynamics of an economic index, Nature,
{\bf 376}, pp. 46-49 (1995)

\medskip 

\noindent [35] Y. Yura, H. Takayasu,
D.  Sornette
and M. Takayasu,   Financial brownian particle in the layered
order-book fluid and fluctuation-dissipation relations,
Physical Review
Letters, {\bf  112}, 098703 (2014)

\medskip

\noindent [36] The Atlas of Economic Complexity: Mapping Paths to 
Prosperity, MIT Press, Cambridge, Massachusetts (2013) 

\noindent [http://atlas.cid.harvard.edu/]

\noindent [https://growthlab.cid.harvard.edu/publications/atlas-economic-complexity-mapping-paths-prosperity]

\medskip

\noindent [37] A Tacchella, M.  Cristelli,
G. Caldarelli, A.  Gabrielli and L. Pietronero,  A new metrics
for countries' fitness and products' complexity. Scientific Reports, {\bf 2},
723 (2012)

\medskip

\noindent [38] L. M. Bettencourt, J.  Lobo, D. Helbing,
C.  Kuhnert and G. B. West,  Growth, innovation, scaling, and
the pace of life in cities. Proceedings of the National Academy of
Sciences, {\bf 104}, pp. 7301-7306 (2007)

\medskip 

\noindent [39] M. R. Frank, L.  Sun, M. Cebrian,
H. Youn and I. Rahwan,   Small cities face greater impact from
automation,  Journal of the Royal Society Interface, {\bf 15}(139),
20170946 (2018)

\medskip

\noindent [40] German Science Foundation `Physics of Socio-Economic
Systems':

\noindent [https://www.dpg-physik.de/dpg/gliederung/fv/soe/index.html]

\noindent [https://www.dpg-physik.de/dpg/gliederung/fv/soe/historie/phbl.html]

\end{document}